\documentclass[aps, prl, twocolumn]{revtex4}
\usepackage{graphicx}
\usepackage{float}
\usepackage{amsmath}
\usepackage{amsfonts}
\usepackage{color}

\begin{document}

\title{Magnetic-Field-Dependent Thermodynamic Properties of Square and Quadrupolar Artificial Spin Ice}

\author{M. Goryca$^{1,2}$, X. Zhang$^3$, J. D. Watts$^{4,5}$, C. Nisoli$^6$, C. Leighton$^4$, P. Schiffer$^{3,7}$, S. A. Crooker$^1$}
\affiliation{$^{1}$National High Magnetic Field Lab, Los Alamos National Laboratory, Los Alamos, NM 87545, USA}
\affiliation{$^{2}$Institute of Experimental Physics, Faculty of Physics, University of Warsaw, Pasteura 5, 02-093 Warsaw, Poland}
\affiliation{$^{3}$Department of Applied Physics, Yale University, New Haven, CT 06520, USA}
\affiliation{$^{4}$Department of Chemical Engineering and Materials Science, University of Minnesota, Minneapolis, MN 55455, USA}
\affiliation{$^{5}$School of Physics and Astronomy, University of Minnesota, Minneapolis, MN 55455, USA}
\affiliation{$^{6}$Theoretical Division, Los Alamos National Laboratory, Los Alamos, NM 87545, USA}
\affiliation{$^{7}$Department of Physics, Yale University, New Haven, CT 06520, USA}

\begin{abstract}
Applied magnetic fields are an important tuning parameter for artificial spin ice (ASI) systems, as they can drive phase transitions between different magnetic ground states, or tune through regimes with high populations of emergent magnetic excitations (\textit{e.g.}, monopole-like quasiparticles).  Here, using simulations supported by experiments, we investigate the thermodynamic properties and magnetic phases of square and quadrupolar ASI as a function of applied in-plane magnetic fields.  Monte Carlo simulations are used to generate field-dependent maps of the magnetization, the magnetic specific heat, the thermodynamic magnetization fluctuations, and the magnetic order parameters, all under equilibrium conditions.  These maps reveal the diversity of magnetic orderings and the phase transitions that occur in different regions of the phase diagrams of these ASIs, and are experimentally supported by magneto-optical measurements of the equilibrium ``magnetization noise'' in thermally-active ASIs. 
\end{abstract}

\maketitle

\section{I. Introduction}

Artificial spin ice (ASI) systems are two-dimensional arrays of lithographically-defined nanomagnets, wherein the interactions between individual elements can be engineered by the size, spacing, and overall geometry of the lattice \cite{Skjaervo:2020, Bramwell:2020, Rougemaille:2019}.  Originally conceived to emulate the frustrated magnetic interactions in  pyrochlore spin ice materials such as Dy$_2$Ti$_2$O$_7$, the essentially unlimited freedom to design ASI lattices has more recently allowed explorations of novel magnetic topologies not found in nature, and in which the degrees of magnetic frustration, extensive degeneracy, and residual entropy can be intentionally engineered \cite{Nisoli:2013, Heyderman:2013}.  

To date, the static and dynamic magnetic properties of ASIs have typically been investigated in zero applied magnetic field. A primary reason for this is that several of the most incisive magnetic imaging tools -- for example magnetic force microscopy (MFM) or photoemission electron microscopy (PEEM) -- work best in the absence of external fields. While transient applied fields are often used in demagnetization protocols that help bring the ASI to its thermodynamic ground state \cite{Wang:2006, Wang:2007, Nisoli:2007, Ke:2008, Nisoli:2010}, or are used to establish an initial out-of-equilibrium magnetic configuration (whose relaxation is subsequently monitored in zero field \cite{Farhan:2013PRL, Farhan:2013}), most experimental and theoretical ASI studies have been principally concerned with their geometry- or temperature-dependent magnetic properties in zero field.  However, it is widely appreciated that an applied magnetic field \textbf{B} can provide an extremely important and versatile tuning parameter in ASIs \cite{Ladak:2010, Mengotti:2011, Phatak:2011, Pollard:2012, Kapaklis:2012, Zeissler:2013, Morley:2019}. Depending on its magnitude and direction, \textbf{B} can be used to drive phase transitions between different magnetic order parameters, or tune through equilibrium regimes where magnetic quasiparticle excitations (e.g., ``magnetic monopoles'') can readily form. 

Our groups have recently studied field-induced thermodynamic effects in two different ASI geometries. In quadrupolar spin ice, we demonstrated the existence of a field-driven phase transition between ferro- and antiferro-quadrupolar  order \cite{Sklenar:2019}, and in conventional square ASI we demonstrated a field-induced monopole plasma regime \cite{Goryca:2021}.  Here we extend and complement those initial studies through additional investigations of the \textbf{B}-dependent thermodynamic properties of these ASI structures.     

We first review the results for square ASI with additional detail beyond the original work \cite{Goryca:2021}, and then we probe both degenerate square ice and the quadrupolar system.  Using Monte Carlo (MC) simulations, we map out the equilibrium magnetization, specific heat, thermodynamic fluctuations, {and magnetic order parameters} of these ASIs as a function of applied in-plane magnetic fields $B_x$ and $B_y$. The calculated maps point to the rich diversity of magnetic orderings and phase transitions that can occur in ASIs in specific regions of their field-dependent phase diagrams.  As in the previous work \cite{Goryca:2021}, we demonstrate agreement with experimental magneto-optical studies of the ``magnetization noise'' in quadrupolar ASI.  Furthermore, the MC calculations allow us to probe the stability of the ferro- and antiferro-quadrupolar phases as  a function of applied  field and temperature; the latter phase is shown to be comparatively fragile, owing to its origin in weaker next-to-next neighbor coupling.  Analogous to the monopole-like quasiparticles that exist in conventional square ASI, quadrupolar ASI is also shown to host regimes of mobile and topologically-protected magnetic excitations at the boundaries between stable magnetic  phases. These results provide a window into a rich landscape of collective magnetic behavior associated with the application of magnetic field to ASI systems.

\section{II. Experimental and theoretical approach}

Our ASI lattices were fabricated from ferromagnetic permalloy (Ni$_{0.8}$Fe$_{0.2}$), following methods reported previously \cite{Gilbert:2016}. Each lithographically-defined nano-island can be approximated as a single Ising-like macrospin with orientation either parallel or antiparallel to the island's long axis. Importantly, the permalloy thickness is engineered to be sufficiently small ($\approx$3.5~nm) that the islands behave as \textit{thermally active superparamagnets} at room temperature. That is, their magnetization direction thermally fluctuates in the absence of a strong biasing magnetic field. These fluctuations ensure that the ASI lattice efficiently samples the vast manifold of possible moment configurations, and remains at or near its thermodynamic ground state and in thermal equilibrium. 

\begin{figure*} 
\center
\includegraphics[width=.99\textwidth]{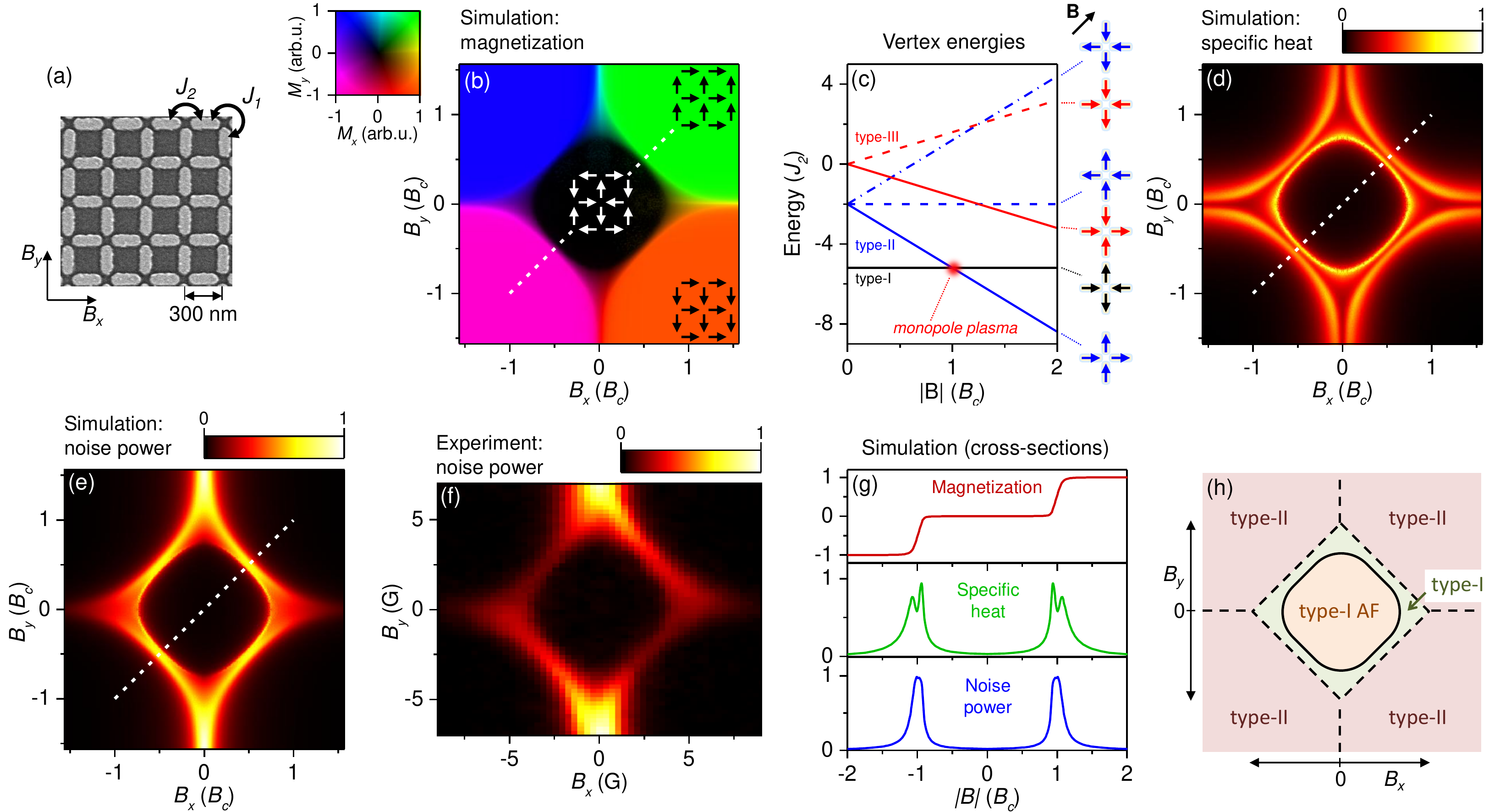}
\caption{(a) SEM image of conventional square ASI. Lateral island dimensions are 220 nm $\times$ 80~nm. Monte Carlo (MC) simulations were used to model its equilibrium thermodynamic properties as a function of applied in-plane magnetic fields $B_x$ and $B_y$, considering only the interaction between nearest (adjacent orthogonal) and next-nearest (adjacent parallel) neighbors, $J_1$ and $J_2$ respectively. Simulations in this figure used $J_1 = 1.8 J_2$, and temperature $kT = 1.2 J_2$. (b) Calculated map of the normalized average magnetization \textbf{M}($B_x, B_y$). The color and brightness indicate the direction and magnitude of \textbf{M}.  The central black region where \textbf{M}$\approx$0 corresponds to type-I ``antiferromagnetic'' order (depicted by white arrows). The regions toward the corners of the map correspond to the four orientations of polarized type-II tiling (black arrows). (c) Field-dependent energies of type-I, type-II, and type-III (monopole) vertices in square ASI, for the case where $B_x = B_y$ (\textit{i.e.}, along the 45$^\circ$ dashed  diagonal line). Type-I and type-II vertices become energetically degenerate when $B_x + B_y = B_c$. (d) Calculated map of the magnetic specific heat $C_m$. (e) Calculated map of the thermodynamic magnetization fluctuations along the $\hat{x}$ axis, $\langle [\delta M_x(t)]^2\rangle$. The diamond-shaped feature indicating large noise reveals the boundary between type-I and type-II tilings, and arises from the kinetics of type-III monopole vertices (see text). (f) Experimentally-measured map of the frequency-integrated magnetization noise power along $\hat{x}$, acquired at $-10^\circ$C, showing good agreement with the simulations. {Data from Ref. \cite{Goryca:2021}.} (g) Cross-sections of the  calculated magnetization, specific heat, and total noise power along the 45$^\circ$ diagonal where $B_x = B_y$. Sharp peaks in $C_m$ indicate phase transitions, while broader peaks typically reveal Schottky anomalies that arise when the field-dependent energies between different vertex types are commensurate with $kT$. (h) Field-dependent phase diagram of conventional square ASI, showing type-I tiling with long-range AF order at small fields, type-I tiling without long-range order at intermediate fields, and polarized type-II tiling at large fields. Monopole regimes emerge along the dashed lines.}
\label{fig_square}
\end{figure*}

The field-dependent thermodynamic properties of these ASI lattices were calculated via standard Glauber MC simulations, which considered nearest, next-nearest, and (where necessary) next-to-next nearest neighbor interactions.  These interaction parameters $J_i$ are defined below, and are chosen to have ratios with each other and with the thermal energy $kT$ that correspond to the actual systems studied. The MC simulations used lattices with 32$\times$32 elements, periodic boundary condictions, and considered single-spin updates only (no cluster flips).  Spins were chosen randomly, and were flipped with probability $p= (1+e^{\Delta/kT})^{-1}$, where $\Delta$ is the energy difference resulting from a spin flip. At each value of $(B_x, B_y)$, approximately 10$^5$ annealing steps were performed, and then the calculated magnetization was recorded for $\sim 10^6$ MC time steps. The average magnetization \textbf{M}, the magnetic specific heat $C_m$ ($\propto \langle [\delta E(t)]^2 \rangle$, where $E$ is the energy of the system), and the thermodynamic magnetization fluctuations $\langle [\delta \textbf{M}(t)]^2 \rangle$ were determined from the computed time series. Additional details of the MC simulations are given in the Appendix. 

For direct comparison with MC simulations, the intrinsic magnetization fluctuations in each ASI lattice were experimentally measured using a recently-developed broadband magneto-optical noise spectrometer \cite{Goryca:2021, Balk:2018}. This ``magnetization noise'' was detected under equilibrium conditions over a wide frequency range (typically 250 Hz to 250 kHz) and was spectrally integrated to give a total noise power that can be directly compared with MC simulations.

\section{III. Conventional Square lattice}

To introduce our approach we first discuss results for the archetypal square ASI lattice \cite{Wang:2006, Moller:2006}, shown in Fig. \ref{fig_square}(a). The essential phenomenology of conventional square ASI can be understood by considering only the coupling $J_1$ between orthogonal nearest-neighbor islands, and the weaker coupling $J_2$ between parallel next-nearest-neighbor islands, as depicted. Figure \ref{fig_square}(b) shows a map of the calculated equilibrium magnetization \textbf{M} versus $B_x$ and $B_y$, where the color and brightness indicate the direction and magnitude of \textbf{M}. This map can be intuitively understood by considering the field-dependent energies of different magnetic configurations at the vertices of the square lattice.  The lowest energy configurations  near zero applied field are ``type-I'' vertices that have 2-in/2-out moment configuration and no net magnetization. In this low-field regime, which corresponds to the central black region of the map where \textbf{M}$\approx$0, the stable ground state is an ordered tiling of type-I vertices, as depicted.  In contrast, if both $|B_x|$ and $|B_y|$ are sufficiently large, then one of the four possible ``type-II'' vertices becomes lowest in energy. Type-II vertices also have 2-in/2-out configuration, but possess a net magnetization along one of the four lattice diagonals.  In these regimes, which correspond to the four colored regions of Fig. \ref{fig_square}(b), the magnetic ground state is simply a fully-polarized tiling of type-II vertices, with stable saturated magnetization oriented along the diagonal that most closely aligns with \textbf{B}.  

Crucially, type-I and type-II vertices, which have energies $\epsilon_{\rm I} = -4J_1 + 2J_2$ and $\epsilon_{\rm II} = -2J_2 - \mu (B_x + B_y)$, must become energetically degenerate at some intermediate field.  As shown previously \cite{Goryca:2021}, this crossover field is given by $|B_x|+|B_y| = B_c = 4(J_1-J_2)/\mu$, where $\mu$ is the magnetic moment of a single nanoisland. This defines the diamond-like shape of the central region where \textbf{M}$\approx$0, and the size of this region is given by the difference between $J_1$ and $J_2$. This intuitive picture highlights the utility of a field-dependent magnetization map for revealing different stable phases of the system, which -- as shown later for the case of quadrupolar ASI -- is especially helpful when considering more complex ASI geometries.

Figure \ref{fig_square}(c) plots the energies of the various types of vertices in conventional square ASI, for \textbf{B} applied along the 45$^\circ$ diagonal indicated by the dashed line in Fig. \ref{fig_square}(b).  As discussed above, type-II vertices (with the appropriate orientation) become the lowest-energy vertex type at the crossover field $B_c$. Also shown are the energies of the type-III vertices, which have unbalanced 3-in/1-out or 3-out/1-in arrangement and therefore possess an effective magnetic charge.  As such, they can be considered as ``monopole-like'' quasiparticle excitations of the square ASI lattice, whose static and dynamics properties have been studied in detail in many earlier works \cite{Ladak:2010, Mengotti:2011, Phatak:2011, Morgan:2011, Pollard:2012, Silva:2012, Kapaklis:2014, Vedmedenko:2016, Morley:2019, Mol:2009}, and whose crucial role in the equilibrium thermodynamics of square ASI will be discussed shortly.

Going beyond previous work \cite{Goryca:2021}, we also calculate the magnetic specific heat $C_m$ from our simulations, which provides an incisive diagnostic because divergences of $C_m$ are typically linked to phase transitions and the emergence of new magnetic order parameters. Figure \ref{fig_square}(d) shows a field-dependent map of $C_m$ in conventional square ASI. The central low-field region is enclosed by a boundary exhibiting large $C_m$, outside of which additional weaker maxima  occur. A cross-section of $C_m$ along the 45$^\circ$ diagonal is shown in Fig. \ref{fig_square}(g); it shows two distinct maxima, on either side of the crossover field $B_c$.  

The sharp maximum of $C_m$ corresponds to the known phase transition to long-range type-I ``antiferromagnetic'' (AF) magnetic order \cite{Morgan:2011, Sendetskyi:2019, Silva:2012, Levis:2013, Moller:2006}.  Importantly, at non-zero temperatures, this  phase transition occurs at applied field $\textbf{B}_{\rm AF}$ that is smaller in magnitude than $B_c$. In the narrow region between $|\textbf{B}_{\rm AF}|$ and $B_c$, type-I vertices are still energetically favored, but long-range AF order is not established.  Precisely at $B_c$, $C_m$ exhibits a local minimum because the type-I and type-II vertex energies are degenerate.  At slightly larger applied fields, however, $C_m$ exhibits a broader second maximum, due to the Schottky-type anomaly that arises when the energy difference between type-I and type-II vertices is comparable to the thermal energy $kT$ (and therefore, large changes of entropy occur for small changes in temperature). Field-dependent maps of $C_m$ therefore provide a powerful tool to identify potential phase transitions to ordered magnetic states, and to help pinpoint degeneracies occurring within the field-dependent manifold of energy levels in ASI. 

Our simulations also give us the intrinsic magnetization fluctuations $\delta \textbf{M}(t)$ in thermal equilibrium, which are also experimentally measurable through the magneto-optical Kerr effect \cite{Goryca:2021}.  Figure \ref{fig_square}(e) shows a map of the calculated power of thermal fluctuations along the $\hat{x}$ direction, $\langle [\delta M_x(t)]^2 \rangle$. The map is characterized by narrow regions of significant noise along the diamond-shaped boundary that separates type-I and type-II tiling (the noise map lacks the four-fold symmetry of the square lattice, because only fluctuations along $\hat{x}$ are shown; a map of $\langle [\delta M_y(t)]^2 \rangle$ is identical, but rotated 90$^\circ$). Regimes of magnetization noise in square ASI are intimately linked to the proliferation and kinetics of type-III monopole vertices \cite{Goryca:2021}.  Within the central dark region of the noise map, fluctuations are suppressed because the system exhibits stable AF type-I magnetic order. In the four corners of the map where both $|B_x|$ and $|B_y|$ are large, all moments are pinned by \textbf{B} and the system exhibits stable type-II tiling. However, crossing over between these regimes near $B_c$ requires flipping individual nano-islands, which necessarily creates type-III monopole vertices. Once created, these monopole excitations can readily diffuse along a staggered diagonal direction. This motion flips spins (causing fluctuations) and converts type-I to type-II vertices (and vice-versa). This process costs no energy at $B_c$ because type-I and type-II vertices are energetically degenerate (see Fig. \ref{fig_square}c). As also shown previously in Ref. \cite{Goryca:2021}, our MC simulations are validated by direct experimental measurements of magnetization noise, shown in Fig. 1(f). 

\begin{figure} 
\center
\includegraphics[width=.49\textwidth]{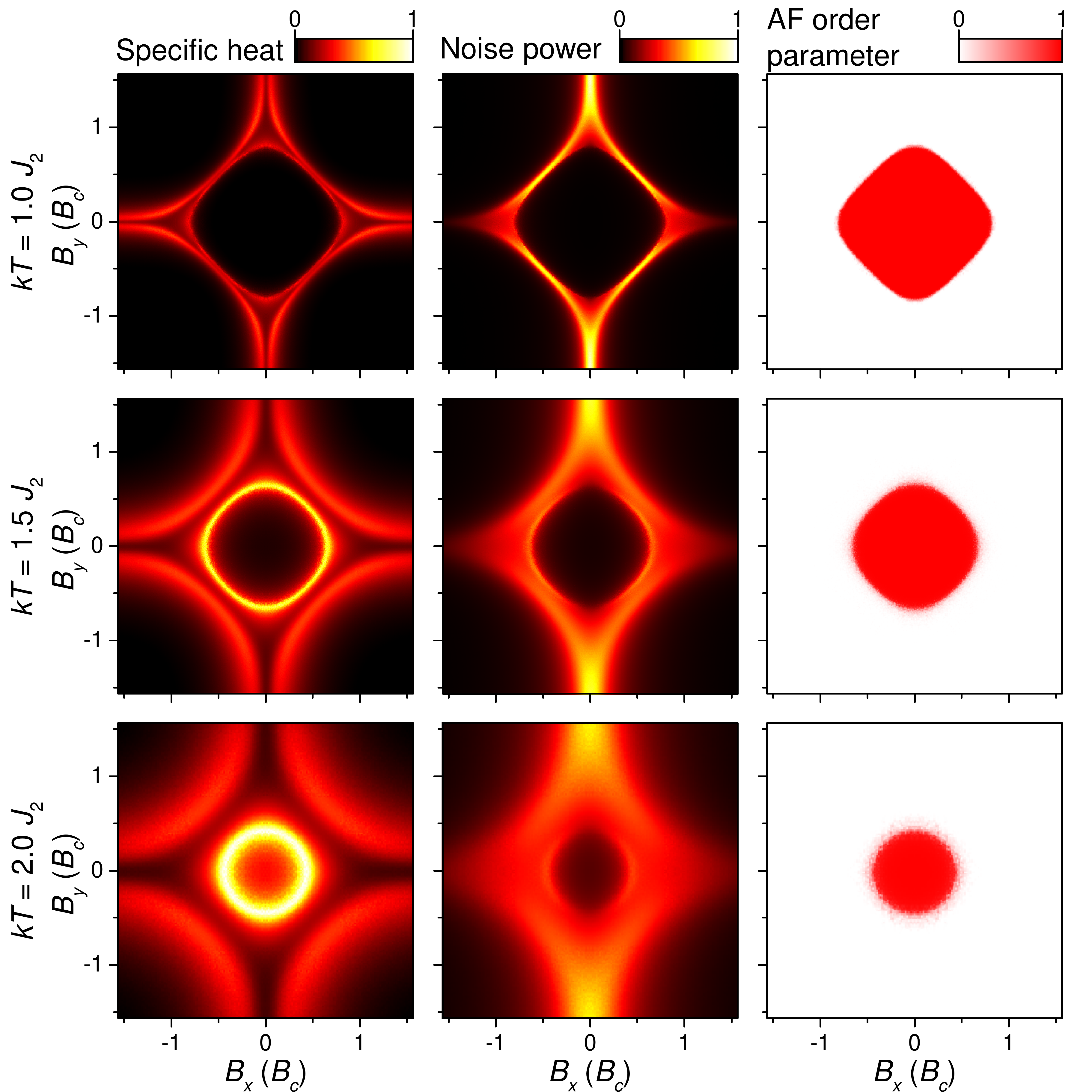}
\caption{Effect of temperature on the field-dependent thermodynamic properties of conventional square ASI.  The left, middle, and right columns show calculated maps of $C_m$, noise power, and the long-range type-I ``antiferromagnetic'' (AF) order parameter {(defined in Appendix)}, respectively.  The top, middle, and bottom row are calculated using $kT/J_2$=1.0, 1.5, and 2.0, respectively.  Field-dependent features in $C_m$ and noise blur with increasing temperature, and the low-field region exhibiting long-range type-I AF order shrinks in size.}
\label{fig2}
\end{figure}

Figure \ref{fig_square}(g) directly compares the magnetization, specific heat, and fluctuations along the 45$^\circ$ diagonal line where $B_x = B_y$. By the fluctuation-dissipation theorem, the frequency-dependent power spectrum of thermodynamic magnetization fluctuations, $S(\omega)$, is linked to the dissipative (imaginary) part of the ac magnetic susceptibility $\chi''(\omega)$ [namely, $S(\omega) \sim kT \chi''(\omega)/\omega$]. Using Kramers-Kronig relations for the zero-frequency dc magnetic susceptibility, $\chi(\omega \sim 0) = (2/\pi) \int d\omega' \chi''(\omega')/\omega'$, we note that the dc susceptibility $\chi = dM/dB \propto \int S(\omega) d\omega$, which is just the total (frequency-integrated) noise power calculated and measured in Figs. 1(e,f).  Figure \ref{fig_square}(g) confirms that the noise is indeed peaked at the crossover field $B_c$ where \textbf{M}(B) varies most rapidly. 

The temperature-dependent properties of ASIs at \textbf{B}=0 have been explored in many prior studies \cite{Budrikis:2012, Kapaklis:2012, Farhan:2013PRL, Levis:2013, Porro:2013, Anghinolfi:2015, Morley:2017, Sendetskyi:2019, Chen:2019, Morley:2019, Pohlit:2020}. As such, Fig. 2 highlights how temperature affects the full \textbf{B}-dependent maps of $C_m$, noise, and long-range magnetic order. As anticipated, the boundaries between different ordered magnetic tilings become less sharp with increasing temperature. Moreover, \textbf{B}$_{AF}$ and therefore the region exhibiting long-range type-I AF order shrinks in size, as shown explicitly by the calculated maps of the AF order parameter. We emphasize, however, that $B_c$ and the location of the monopole-rich regime are unaffected by temperature, since $B_c$ depends only on the difference $J_1 - J_2$.

Interestingly, the region where AF order appears not only shrinks as temperature increases, but changes shape (see Fig. \ref{fig2}). We can understand this phenomenon in the context of approaching the effective ordering temperature, \emph{i.e.}, the N{\'e}el temperature for the ordered type-I phase. At zero temperature, the AF region must be diamond shaped, because its boundary must coincide with the degeneracy line between type-I and type-II vertices. This is dictated by considerations of energy balance, and the boundary shape reflects the anisotropy of the square lattice. As the temperature increases, however,
the anisotropy of the diamond-shaped AF region is lost and its shape becomes more circular. A possible explanation can be found in the nature of critical behavior and scale-invariance at a second order phase transition \cite{Kadanoff:1966}. As the type-I ordered phase approaches criticality at the N{\'e}el temperature, the correlation length of fluctuations grows to be much larger than the lattice constant, and the order parameter fluctuations are no longer impacted by the underlying square lattice structure. As a result, the field anisotropy to the order parameter is eliminated.

\begin{figure*} 
\center
\includegraphics[width=.8\textwidth]{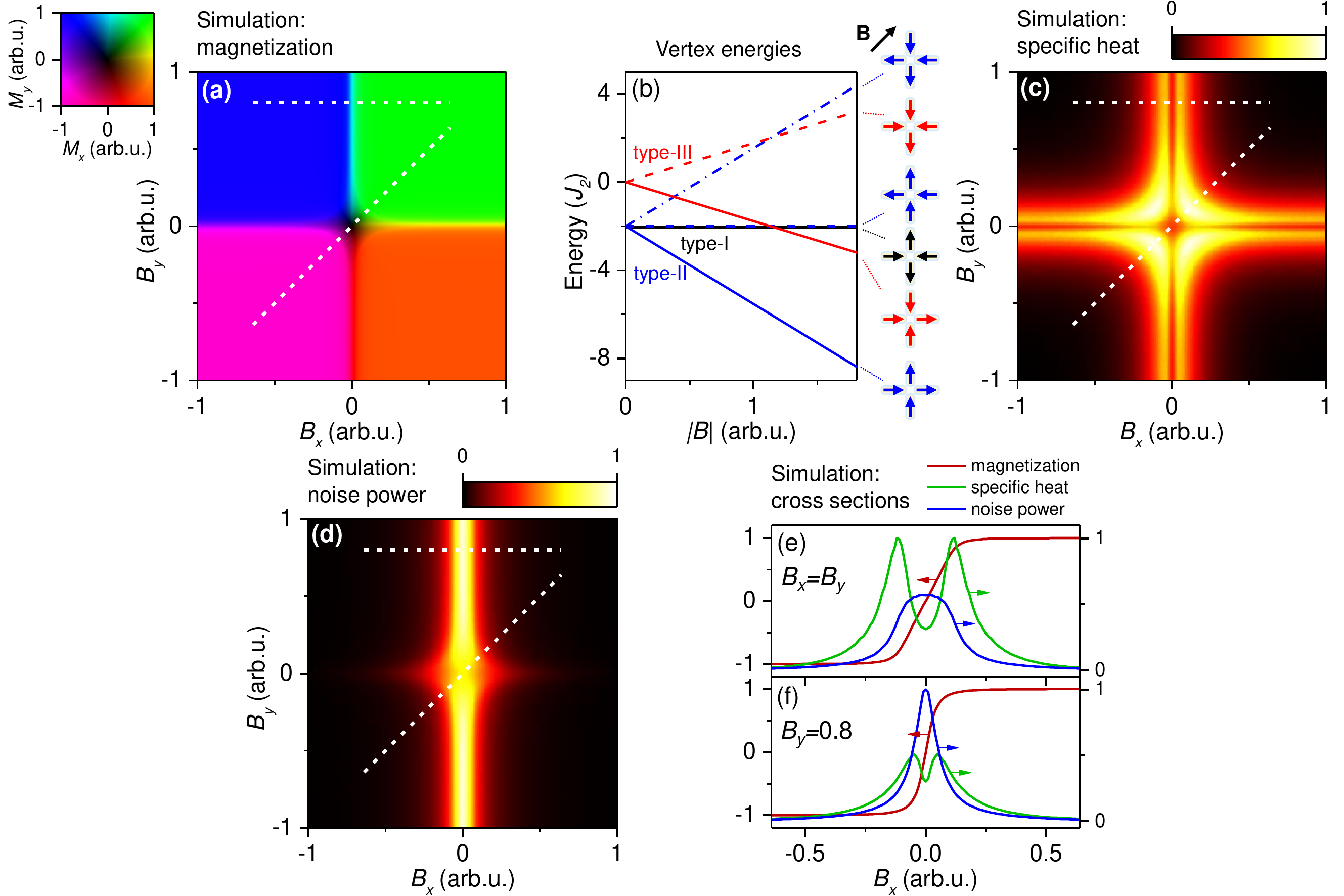}
\caption{MC simulations of ``degenerate'' square ASI, where $J_1 = J_2$ and $kT = 1.2 J_2$. Here, type-I and type-II vertices have equal energy at zero applied field. (a) Calculated map of the field-dependent average magnetization, \textbf{M}. (b) Field-dependent energies of type-I, type-II, and type-III (monopole) vertices, along the 45$^\circ$ diagonal line where $B_x = B_y$ [\textit{cf.} Fig. 1(c)]. (c) Calculated map of the magnetic specific heat, $C_m$. (d) Calculated map of the  thermodynamic magnetization noise along the $\hat{x}$ axis. (e) Cross sections of \textbf{M}, $C_m$, and noise along the 45$^\circ$ diagonal where $B_x = B_y$ (red, green, and blue lines, respectively). (f) Cross sections of \textbf{M}, $C_m$, and noise along the horizontal dashed line where $B_y=0.8$.}
\label{fig_square_deg} 
\end{figure*}

Taken together, these simulations and experiments allow us to construct the field-dependent phase diagram of conventional square ASI, shown in Fig. 1(h). Specific heat identifies the phase boundaries denoting the onset of long-range type-I order (solid line). Magnetization and noise maps help to pinpoint the crossover between type-I and type-II tiling, and also between type-II tilings with different orientation, (dashed lines), where a high density of magnetic monopoles energe. The case of conventional square ASI highlights the utility of this methodology, which we now apply to degenerate square ASI and to the recently-introduced quadrupolar ASI lattice.

\section{IV. Degenerate square ASI}

A special case of square ASI that attracts much attention is that of ``degenerate square ice'', which occurs when the lattice  is designed so that $J_1 \approx J_2$ (for example, by engineering a height offset between the horizontal and vertical islands, as demonstrated in pioneering recent experiments by Perrin \textit{et al.} \cite{Perrin:2016} and Farhan \textit{et al.} \cite{Farhan:2019}).  In this case, the energies of type-I and type-II vertices are equal at \textbf{B}=0. The lowest-energy magnetic configuration at \textbf{B}=0 is therefore a disordered mix of the six possible type-I and type-II vertices, with concomitant extensive degeneracy and large residual entropy. As such, degenerate square ice approximates the magnetic interactions in pyrochlore spin ice materials such as the rare-earth titanates Dy$_2$Ti$_2$O$_7$ and Ho$_2$Ti$_2$O$_7$ \cite{Harris:1997, Castelnovo:2008, Bramwell:2009, Dusad:2019}, including the thermal creation, annihilation, and motion of  monopole-like quasiparticle excitations (\textit{i.e.}, type-III vertices).

Following our approach for conventional square ice, we used MC calculations with $J_1 = J_2$, and $kT =1.2J_2$. Figure \ref{fig_square_deg} shows the resulting field-dependent maps of the thermodynamic properties.  In line with expectations, the map of \textbf{M}$(B_x, B_y)$ is similar to that for conventional square ASI (\textit{cf.} Fig. 1b), but with the central region (where stable type-I order exists when $J_1>J_2$) shrunk to zero size.  Stated differently, the crossover field where type-I and type-II vertices have equal energy, $B_c = 4(J_1 - J_2)/\mu$, now occurs at zero field [see also the field-dependent vertex energies shown in Fig. \ref{fig_square_deg}(b)].  Application of both $|B_x|>0$ and $|B_y|>0$ uniquely lifts this degeneracy, and favors the type-II vertex with moment most closely aligned with \textbf{B}, leading to a saturated \textbf{M}. A cross-section through this map is shown in Fig. \ref{fig_square_deg}(e), for \textbf{B} applied along a 45$^\circ$ diagonal.  The characteristic field scale at which \textbf{M} saturates is given by the ratio of $kT/J_2$. 

A map of the specific heat $C_m$ [Fig. \ref{fig_square_deg}(c)] shows only broad features associated with Schottky anomalies that occur when the difference between vertex energies, $\epsilon_{\rm I} - \epsilon_{\rm II}$, is commensurate with $kT$. No sharp or divergent features are observed on this map, in accordance with the expectation that degenerate square ASI does not exhibit any phase transitions to an ordered magnetic state. 

The calculated map of the thermodynamic noise power, $\langle [\delta M_x(t)]^2 \rangle$, indicates significant fluctuations only along the vertical stripe where $B_x \approx 0$ [see Fig. \ref{fig_square_deg}(d)]. Similar to \textbf{M} and $C_m$, this map is reminiscient of that from conventional square ice, but with the central region shrunk to zero size.  Fluctuations at zero field in degenerate square ice arise from the thermal creation, annihilation, and motion of type-III monopole vertices. Once created, these monopoles can diffuse within the lattice without cost in energy at \textbf{B}=0, leading to fluctuations. Note, however, that monopole diffusion in degenerate square ice at zero field can occur in any direction, whereas monopole diffusion in conventional square ice at $B_c$ occurs primarily along the staggered lattice diagonal that is most closely aligned with \textbf{B}.

Application of non-zero $B_{x}$ and $B_y$ induces the monopoles to move toward the edges of the lattice, leaving behind a polarized and stable type-II lattice tiling with saturated \textbf{M}, in which fluctuations are suppressed.  Exceptions to this trend occur when the field is applied exactly along the horizontal or vertical direction. In this case, two of the four possible type-II vertices remain degenerate, and the Ising-like moments of the islands that are oriented orthogonal to \textbf{B} remain unbiased and free to fluctuate thermally. Consider the case where $B_x=0$ and $B_y >0$ (\textit{i.e.}, the vertical stripe on the noise map). Here, all vertical islands are polarized, but the horizontal islands do not `feel' any net bias and can still fluctuate thermally, giving large noise.  This process can be regarded as an effective dimensional reduction, where thermally-created type-III monopoles are constrained to diffuse along the 1D chains of horizontal islands. We note that a similar regime of 1D monopole diffusion exists in conventional square ASI when $B_x=0$ and $|B_y| >B_c$.

Figure \ref{fig_square_deg}(e) shows cross sectional plots through the maps of \textbf{M}, $C_m$, and noise along the 45$^\circ$ diagonal line where $B_x = B_y$. These can be compared with the analogous 45$^\circ$ line cuts shown in Fig. 1(g) for conventional square ASI. As in conventional square ASI, $C_m$ achieves a minimum -- and fluctuations are most pronounced -- when the energies of type-I and type-II vertices are degenerate and monopole-like type-III vertices can proliferate and diffuse freely. Also shown [Fig. \ref{fig_square_deg}(f)] are cross sections along the horizontal line cut where $B_y=0.8$. As  discussed in the previous paragraph, along this line all vertical islands are polarized by $B_y$, and the noise power $\langle [\delta M_x(t)]^2 \rangle$ achieves a maximum at $B_x=0$ when the horizontal islands are effectively unbiased and are free to fluctuate.  While we do not have experimental results to compare with the MC simulations in this case, the results show the utility of simulations to predict the magnetic phase diagram of this system, including the regimes densely populated with monopoles.

\begin{figure*} 
\center
\includegraphics[width=.99\textwidth]{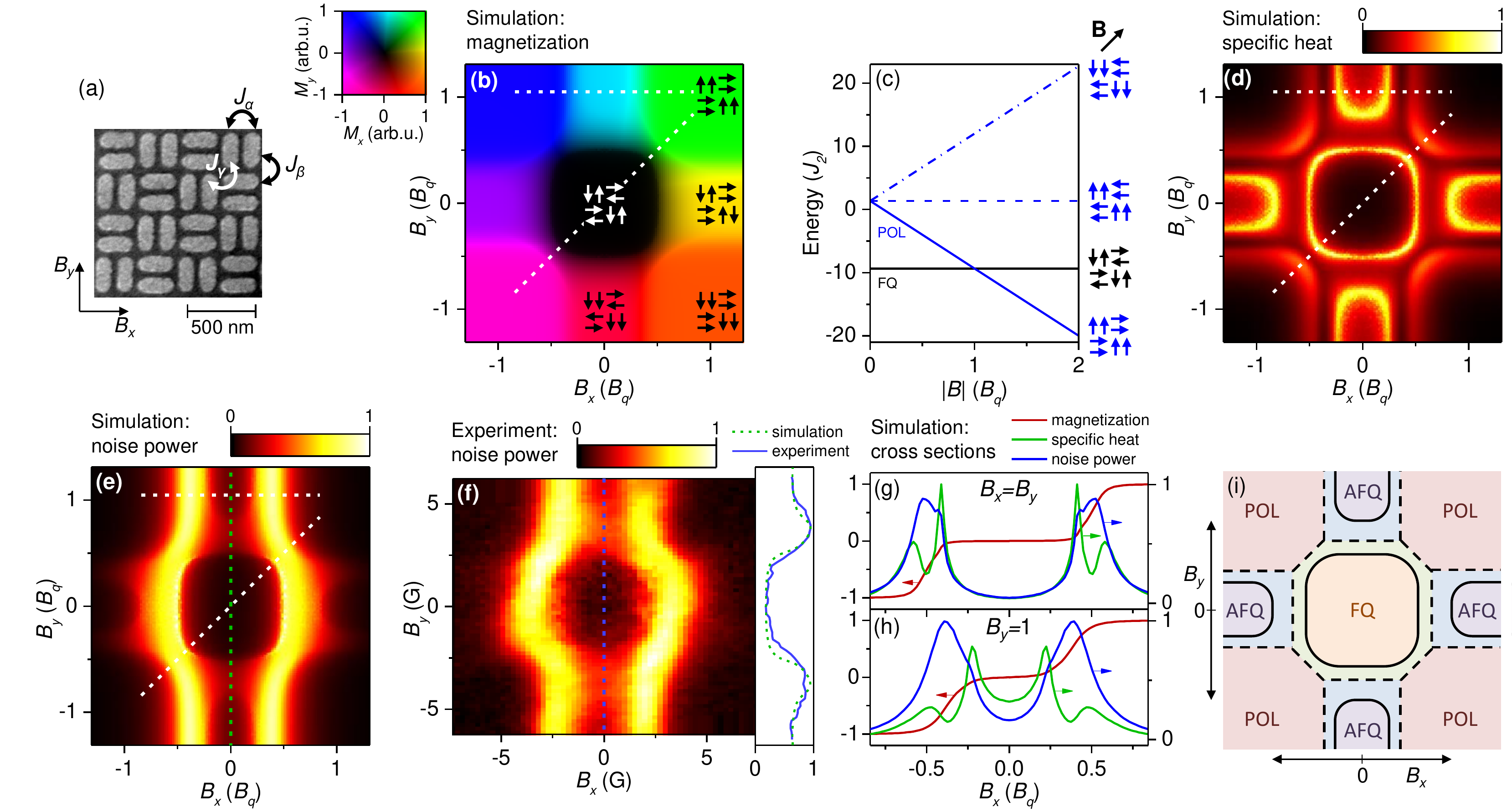}
\caption{(a) SEM image of quadrupolar ASI. Lateral island dimensions are $\approx$ 220 nm $\times$ 100~nm.  MC simulations consider the interactions $J_\alpha$ (between parallel adjacent islands), $J_\beta$ (between perpendicular adjacent islands), and $J_\gamma$ (between parallel next-nearest islands), as depicted.  These simulations use $kT = 1.2 J_\beta$, and $J_\alpha = 2 J_\beta = 3 J_\gamma$ (in approximate accordance with micromagnetic simulations which give $J_\alpha = 2.3 J_\beta = 4.3 J_\gamma$). (b) Calculated map of \textbf{M}($B_x, B_y$). Arrows depict the magnetic ordering in different regions. (c) Energies of different quadrupole moment configurations, versus applied field along a 45$^\circ$ diagonal ($B_x = B_y$). The crossover field between FQ and polarized tiling is $B_q$. (d) Calculated map of the magnetic specific heat, $C_m$. (e) Calculated map of the magnetization noise along $\hat{x}$.  (f) Experimentally-measured map of the magnetization noise along $\hat{x}$, acquired at $+7^\circ$C, showing good agreement. Note the reduced noise at $B_x=0$ and large $|B_y|$, which corresponds to the AFQ phase (the sidebar compares vertical cross sections through both simulated and measured noise). (g,h) Comparing \textbf{M}, $C_m$, and noise along a 45$^\circ$ line cut ($B_x = B_y$), and along a horizontal line cut ($B_y$=1). (i) Field-dependent phase diagram of quadrupolar ASI, showing regimes of polarized tiling, and long-range FQ and AFQ order. { In the green and blue regions, FQ and AFQ tiling is energetically favored (respectively), but \textit{long-range} magnetic order cannot stabilize due to the non-zero temperature.  Crossovers between regimes with different favored tilings are indicated with dashed lines (along which different tilings are energetically degenerate), and are rich in topologically-protected magnetic excitations.}}
\label{fig_quadrupolar}
\end{figure*}

\section{V. Quadrupolar ASI lattice}

The quadrupolar ASI lattice, introduced in 2019 by Sklenar \textit{et al.} \cite{Sklenar:2019} and shown in Fig. \ref{fig_quadrupolar}(a), consists of pairs of parallel nano-islands, arranged horizontally and vertically in a checkerboard pattern. As shown previously \cite{Sklenar:2019}, in the absence of applied magnetic fields, the strong coupling between the two islands within a pair maintains their opposite moment orientation; such a configuration  has no net dipole moment but does possess a quadrupole moment.  Weaker coupling between orthogonal islands in adjacent pairs can lead to long-range ``ferroquadrupolar'' (FQ) order of the entire lattice. That previous study demonstrated that a field aligned with one of the principal axes of the lattice can drive a transition from ferroquadrupolar to antiferroquadrupolar  (AFQ) order, marking the first thermodynamic mapping of a field-dependent phase diagram in ASI, albeit along only a single field direction.

We now expand upon that previous work with the same approach of MC simulations and noise measurements as described above for square ASI. To avoid confusion with initial studies \cite{Sklenar:2019}, which considered the pairs as fundamental units and used ``$J_{1,2}$'' to denote couplings \textit{between} different pair units, here we use the parameter $J_\alpha$ to indicate the coupling between an individual island and its parallel neighbor within a pair, and $J_\beta$ to indicate the weaker coupling between an island and each of the four orthogonal islands in adjacent pairs, as indicated in Fig. \ref{fig_quadrupolar}(a). We also consider the still-weaker coupling $J_\gamma$ between parallel islands in next-nearest neighbor pairs. $J_\gamma$ permits the emergence of long-range AFQ ordering at certain applied fields. 

Figure \ref{fig_quadrupolar}(b) shows the calculated map of \textbf{M}($B_x, B_y$).  The map exhibits a low-field region where $\textbf{M} \approx 0$, consistent with stable FQ magnetic order wherein islands within every pair are oriented oppositely (due to $J_\alpha$), and where neighboring pairs have a specific relative orientation (due to $J_\beta$). However, when both $|B_x|$ and $|B_y|$ are sufficiently large (\textit{i.e.}, in the four corners of the map), both islands within every pair necessarily become polarized along the direction most closely aligned with \textbf{B}, and the lattice exhibits trivial polarized order, as depicted.  Figure \ref{fig_quadrupolar}(c) shows the field-dependent energies of different moment configurations, along the 45$^\circ$ line where $B_x = B_y$. At zero field, FQ tiling is favored, but at a crossover field $B_q$ the energies of FQ and (a specific) polarized moment configuration become degenerate. (Note, however, that a thermodynamic phase transition to \textit{long-range} FQ order occurs precisely at $B_q$ only at zero temperature; analogous to long-range type-I order in conventional square ice, the phase transition to long-range FQ order occurs at fields smaller than $B_q$ when $T>0$.) The crossover field $B_q$ occurs when $|B_x| + |B_y| = B_q = (2J_\alpha + 4 J_\beta - 4 J_\gamma)/\mu$. However, in contrast to the diamond-shaped boundary separating type-I and type-II tiling in conventional square ASI, these diagonal boundaries in quadrupolar ASI are truncated such that the boundary enclosing the low-field FQ tiling region is approximately octagonal.  The vertical (horizontal) sides of this boundary are given by $|B_x| (|B_y|) = (J_\alpha + 4 J_\beta - 4 J_\gamma)/\mu$, respectively.

Returning to the map of \textbf{M}, we note that for the special case of large \textbf{B} applied along (or nearly along) a principal axis where $B_x \approx 0$ or $B_y \approx 0$, pairs of islands oriented parallel to \textbf{B} become polarized, but orthogonally-oriented pairs do not. This leads to relatively wide regions, shown for example by the yellow and aqua colors, where the net magnetization remains relatively constant. We show below that the AFQ ordered phase emerges in these regimes.

The specific heat map shown in Fig. \ref{fig_quadrupolar}(d) exhibits a closed boundary where $C_m$ is large and divergent [see also the cross section shown in Fig. \ref{fig_quadrupolar}(g)], indicating the thermodynamic phase transition to \textit{long-range} FQ order within the enclosed low-field region. Analogous to conventional square ASI, this phase transition occurs at an ordering field less than $B_q$ at non-zero temperatures. Outside this region, additional weak maxima and minima of $C_m$ reveal Schottky anomalies and level crossings, respectively, due to the field-dependent energies of the different moment configurations [as shown, \textit{e.g.}, in Fig. \ref{fig_quadrupolar}(c)]. 

Especially interesting are the four additional U-shaped regions revealed by boundaries of large and divergent specific heat at the edges of the map in Fig. \ref{fig_quadrupolar}(d), where $|B_x|$ is large (and $|B_y|$ is small), or where $|B_y|$ is large (and $|B_x|$ is small).  These boundaries reveal the phase transitions to long-range AFQ order, wherein next-nearest pairs of islands order relative to one another, driven by $J_\gamma$. These phase transitions manifest clearly in $C_m$ and also (to a lesser extent) in the magnetization noise.  

As shown in Fig. \ref{fig_quadrupolar}(e), the calculated noise $\langle [\delta M_x(t)]^2 \rangle$ does exhibit dark regions of low noise, not only in the low-field FQ phase and in the fully-polarized regimes, but also when $|B_y|$ is large and $|B_x| \approx 0$, indicating stable AFQ magnetic order.  This regime is separated from the large central FQ regime by a boundary of additional noise at intermediate $|B_y|$. Crucially, the experimental noise data validate the MC simulations and fully support the presence of stable AFQ ordering in our thermally-active quadrupolar ASI sample. Figure \ref{fig_quadrupolar}(f) shows that the measured noise not only clearly reveals the large region of stable FQ order at small \textbf{B}, but also shows additional regions of low noise at the upper and lower edges of the map where $|B_y| \gg 0$ and $|B_x| \approx 0$, in good agreement with simulations. Both the calculated and measured noise maps along the $B_x =0$ vertical cross-section show clear local minima at large $|B_y|$, consistent with stable AFQ order. These minima are separated from the FQ phase by a local maximum in the magnetization fluctuations where stable long range order does not occur.  

Cross sections of \textbf{M}, $C_m$, and noise, along the diagonal and horizontal line-cuts indicated, are shown in Figs. \ref{fig_quadrupolar}(g) and \ref{fig_quadrupolar}(h).  As before, sharp peaks in $C_m$ demarcate phase transitions to different magnetic order parameters, and noise peaks reveal degeneracies between different magnetic configurations. These data allow us to construct a two-dimensional field-dependent phase diagram for quadrupolar ASI, shown in Fig. \ref{fig_quadrupolar}(i). Regions exhibiting long-range FQ and AFQ order are indicated, as are high-field regions exhibiting trivial polarized order. Separating these phases are narrower regimes where particular magnetic configurations are energetically favored, but where long-range order is not stable due to the non-zero temperature. {As shown below, crossovers between those regimes (dashed lines in Fig. \ref{fig_quadrupolar}(i)), at which different magnetic configurations are energetically degenerate, exhibit high density of magnetic excitations, in analogy to the case of square ASI.}

\begin{figure} 
\center
\includegraphics[width=.49\textwidth]{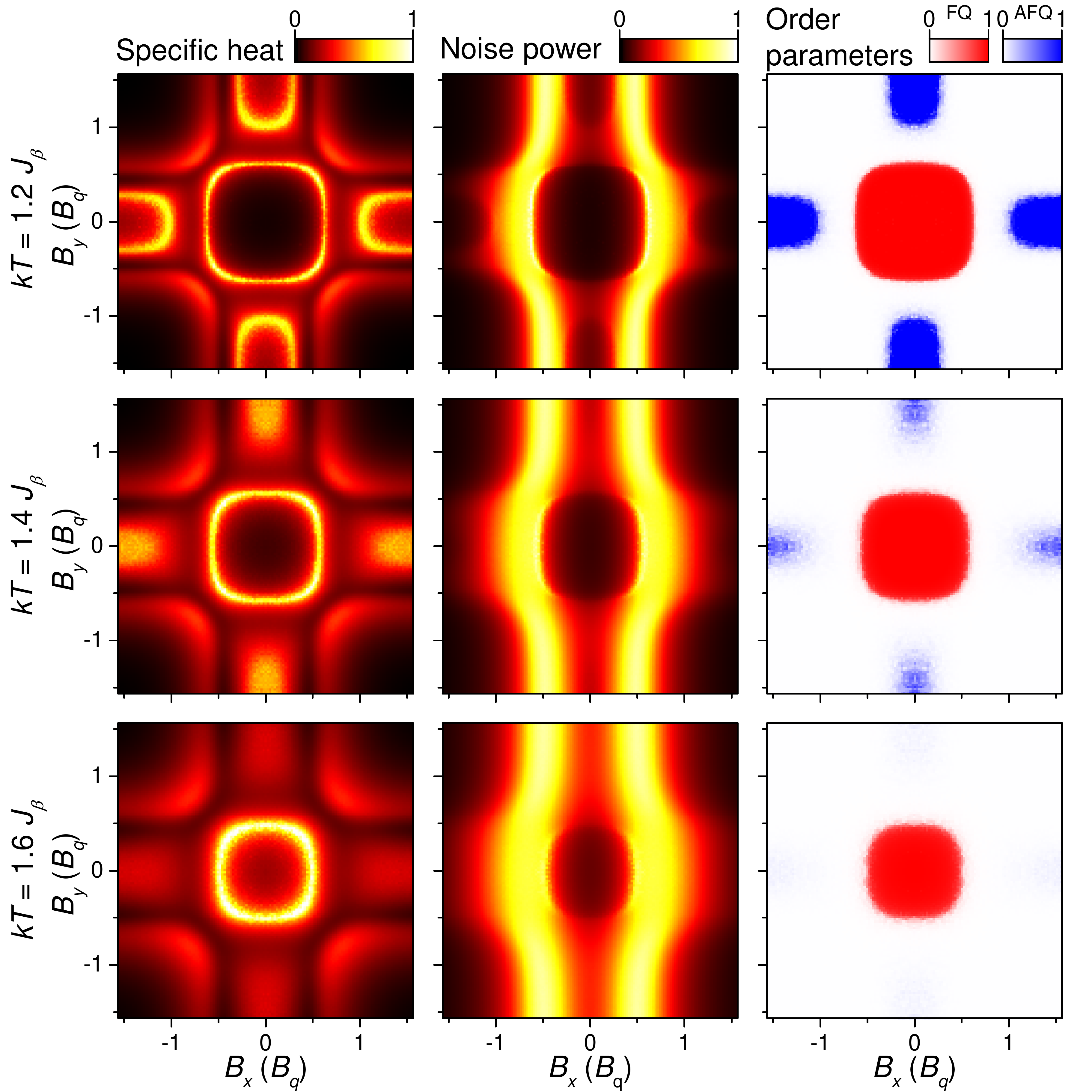}
\caption{Effect of temperature on quadrupolar ASI, showing the collapse of the AFQ phase as temperature $kT$ increases from $1.2 J_\beta$ to $1.4 J_\beta$ to $1.6 J_\beta$ (top, middle, and bottom rows, respectively).  Left, middle, and right columns show maps of $C_m$, noise, and both FQ and AFQ order parameters {(see Appendix)}, respectively.}
\label{quad_temp}
\end{figure}

The effect of temperature, particularly on the AFQ phase, is shown in Fig. \ref{quad_temp}. For $kT > 1.4 J_\beta$, MC simulations show  that long-range AFQ order disappears, whereas the more robust FQ phase remains (albeit diminished in size). At elevated temperature, the boundary of large $C_m$ that formerly surrounded the AFQ phase no longer appears, and the equilibrium noise no longer exhibits a broad local minimum in this regime but is instead uniform.  Calculations of the AFQ and FQ order parameters indicate that long-range AFQ ordering no longer exists by $kT = 1.6 J_\beta$. The size and stability of the AFQ phase depends largely on $J_\gamma$, which is considerably smaller than $J_\alpha$ and $J_\beta$, and therefore the fragile AFQ phase is correspondingly much less stable against increasing temperature. 


Finally, we note similarities between the regions of large magnetization noise that occur near $B_c$ in square ASI [\textit{cf.} Fig. 1] -- which revealed phases rich in magnetic monopoles -- and the regions of large noise that occur near $B_q$ in quadrupole ASI. In both cases, crossing between the two degenerate magnetic configurations (namely, between type-I and type-II vertices at $B_c$ in square ASI, or between FQ and polarized arrangements at $B_q$ in quadrupolar ASI), necessarily requires the creation of pairs of higher-energy magnetic configurations. In square ASI these are the topologically-protected type-III monopole vertices.  In quadrupole ASI, these ``type-C'' excitations are in sets of four moments that are analogous to vertices of the pinwheel lattice \cite{Macedo:2018} and the trident lattice \cite{trident}.  

\begin{figure} 
\center
\includegraphics[width=.45\textwidth]{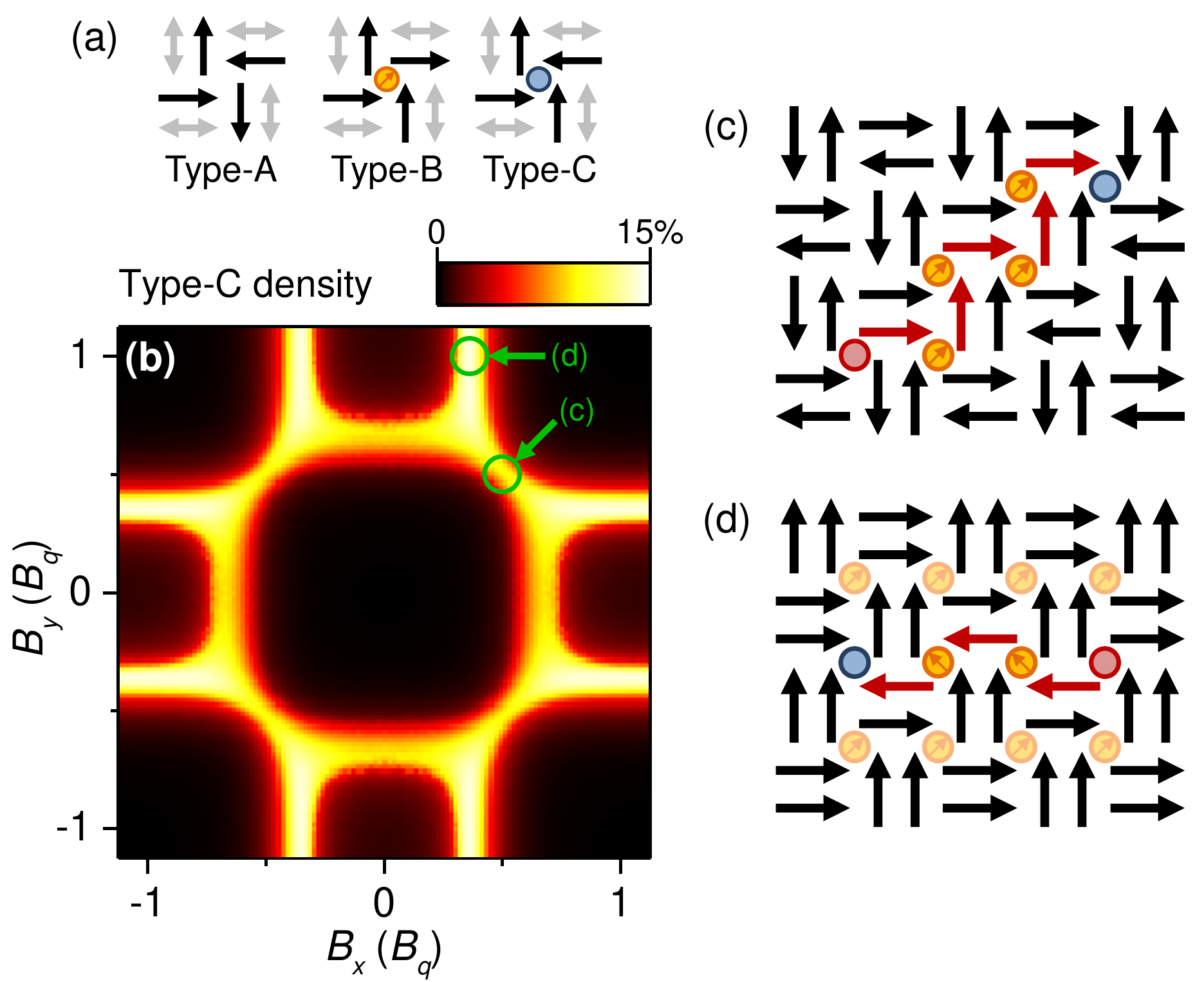}
\caption{(a) Schematic of different moment configurations in quadrupolar ASI (type-A, -B, and -C), analogous to the type-I, -II, and -III vertices in square ASI. Blue/red dots indicate type-C configurations with three moments pointing towards/away from the center, respectively. (b) A map of the equilibrium density of type-C excitations when $kT/J_2 = 1.0$. {Analogously to the case of square ASI (see Fig. 4 in Ref. \cite{Goryca:2021}), regions of high excitation density track crossovers between different magnetic orderings (dashed lines in Fig. \ref{fig_quadrupolar}(i)).} (c) A depiction of how these excitations form and can diffuse, at $(B_x, B_y) = (0.5 B_q, 0.5 B_q)$ (specifically, as FQ tiling switches to polarized tiling, and where type-A and -B configurations are energetically degenerate).  In this case, pairs of type-C excitations diffuse along a staggered diagonal direction. Red arrows denote spins that have flipped. (d) Same, but at $(B_x, B_y) = (0.35 B_q, 1.0 B_q)$ (specifically, as polarized order switches to AFQ order).  In this case the type-C excitations diffuse primarily along a horizontal direction. Switching between ordered tilings of type-A and type-B configuration, or between differently-polarized type-B tilings, requires generation and motion of type-C configurations.}
\label{quad_monopoles}
\end{figure}

Figure \ref{quad_monopoles}(a) depicts the relevant moment configurations in quadrupolar ASI. Type-A and -B configurations have the lowest energy in the FQ and polarized regions, respectively. The higher-energy type-C excitations are generated in pairs, are topological, and can diffuse freely through the lattice when $|\textbf{B}|=B_q$.  Furthermore, as charged pairs of type-C excitations separate, they leave a line of type-A or -B in between, in direct analogy again to the square ice monopole plasma discussed above.

The analogy between square and quadrupolar ASI stems from the fact that they are topologically equivalent, with $J_1$ and $J_2$ in square ASI being equivalent to $J_{\beta}$ and $J_{\gamma}$, respectively. In this mapping, the AF phase in the square lattice corresponds to the FQ tiling in the quadrupolar lattice, whereas AFQ tiling in the quadrupolar lattice corresponds to square ice with (for example) all vertical islands polarized in the same direction, and polarized rows of horizontal islands oriented in alternating directions. Note that in actual realizations of square ASI, the equivalent of $J_{\alpha}$ is much less than $J_1$ and $J_2$, and thus the equivalent of an AFQ phase is never observed. 

Because of this topological equivalence, the kinetic pathways for single-spin dynamics are identical. In other words, just as a single spin flip within type-I or type-II ordered square ASI creates a pair of type-III monopoles that can be further separated by additional spin flips, in quadrupolar ASI a single spin flip within FQ or AFQ tiling creates a pair of type-C excitations. Thus, type-C excitations drive the kinetics in quadrupolar ASI, much as type-III monopole kinetics drive the kinetics in square ASI. Therefore, similar to the case of square ASI, the maximum noise power corresponds to the maximum density of type-C excitations.

Figure \ref{quad_monopoles}(b) shows a field-dependent map of the calculated average density of these excitations in thermal equilibrium.  {As expected,} they exist primarily along the boundaries separating regions of different favored magnetic tiling{, again in analogy to the case of square ASI}. Figures \ref{quad_monopoles}(c,d) illustrate how these excitations can propagate through the quadrupole lattice, at two different points on the magnetic phase diagram.  The first depicts the system at the boundary between FQ and polarized tiling ($B_x = B_y = 0.5 B_q$); in this case the excitations can readily diffuse along a staggered lattice diagonal. The second case depicts the system at the boundary between AFQ and polarized tiling ($B_x = 0.35 B_q, B_y = 1.0 B_q$); here the excitations diffuse primarily along a horizontal direction.  We note that the commonality of the boundary regime of mobile charged excitations in both lattice geometries suggests that this may be a generic feature in artificial spin ices, associated with the transition from an ordered state with no net polarization to a polarized ordering. 

\section{VI. Summary}

In summary, we have investigated the magnetic-field-dependent equilibrium thermodynamic properties (magnetization, specific heat, fluctuations, and magnetic order parameters) in both square ASI (including the degenerate square lattice) and quadrupolar ASI, using both MC simulations and direct experimental measurements of magnetization noise. The \textbf{B}-dependent maps reveal the rich diversity of magnetic orderings and phase transitions that can occur in ASIs in specific regions of their field-dependent phase diagrams, which we are able to reconstruct using these different thermodynamic properties.  Both calculated and measured maps show regions of stable long-range magnetic order, which are typically separated by boundaries where topologically-protected magnetic excitations can readily proliferate and diffuse. We reveal the nature of those excitations in quadrupolar ASI and discuss their similarities to the magnetic monopole quasiparticles that emerge in square ice. These results motivate further exploration of how applied magnetic fields can drive novel magnetic phases in more complex ASI lattice geometries, and especially the possibilities of mobile magnetic charges in systems with different geometrical configurations.

\section{Appendix: MC simulations}
Typical lattices used in our MC simulations had $32 \times 32$ islands in the case of the square ASI, and $32 \times 32$ pairs of islands in the case of the quadrupolar ASI, both with periodic boundary conditions. The energy of the square lattice was calculated as:

\begin{equation*}
\epsilon_{\rm s}=\sum_{i}\left(\sum_{j \in {\rm NN}} \frac{J_1 s_i s_j}{2} + \sum_{j \in {\rm NNN}} \frac{J_2 s_i s_j}{2} + {\rm \textbf{B}}\cdot\boldsymbol{s_i}\right),
\end{equation*}

where $s_i$ is the orientation of the $i$-th spin, that is $s_i = \pm 1$ when the island's moment is parallel/antiparallel to its long axis, while ${\rm NN}$ and ${\rm NNN}$ denote the four nearest neighbors (adjacent orthogonal) and the two next-nearest neighbors (adjacent parallel) of the $i$-th spin, respectively (see also Fig. \ref{fig_square}(a)). 

The energy of the quadrupolar lattice was calculated as:


\begin{equation*}
\begin{split}
\epsilon_{\rm q} = \sum_{i}\left(\sum_{j \in {\rm NN}} \frac{J_{\alpha} s_i s_j}{2} + \sum_{j \in {\rm NNN}} \frac{J_{\beta} s_i s_j}{2} \right. \\\
\left. + \sum_{j \in {\rm NNNN}} \frac{J_{\gamma} s_i s_j}{2} + {\rm \textbf{B}}\cdot\boldsymbol{s_i}\right),
\end{split}
\end{equation*}

where ${\rm NN}$ is the nearest neighbor of the $i$-th spin (\textit{i.e.}, its sole parallel neighbor within a pair), ${\rm NNN}$ are its next-nearest neighbors (the four orthogonal islands in adjacent pairs), and  ${\rm NNNN}$ are its next-next-nearest neighbors (the two parallel islands in next-nearest neighbor pairs -- see also Fig. \ref{fig_quadrupolar}(a)).

The MC simulations used single-spin updates only (no cluster or loop flips).  Spins were chosen randomly, and were flipped with probability $p= (1+e^{\Delta/kT})^{-1}$, where $\Delta$ is the energy difference resulting from a spin flip. At each value of $(B_x, B_y)$, approximately 10$^5$ annealing steps were performed, and then the calculated magnetization was recorded for $\sim$$10^6$ MC time steps. The average magnetization \textbf{M}, the magnetic specific heat $C_m$ ($\propto \langle [\delta E(t)]^2 \rangle$, where $E$ is the energy of the system), and the thermodynamic magnetization fluctuations $\langle [\delta \textbf{M}(t)]^2 \rangle$ were determined from the computed time series. 

For the annealed lattices we also calculated order parameters. For the square lattice, the AF order parameter was defined as: 

\begin{equation*}
\eta_{\mathrm{AF}}=\frac{1}{n}\left|\sum\limits_{x,y}\left(h_{x,y} - v_{x,y}\right)(-1)^{x+y}\right|, 
\end{equation*}

where $h_{x,y}$ and $v_{x,y}$ denote the moment of the horizontal and vertical island at coordinates $(x,y)$ (see Fig. 7a), and $n$ is the total number of spins in the simulated system.
 
For the quadrupolar lattice, the FQ order parameter was defined as:

\begin{equation*}
\eta_{\mathrm{FQ}}=\frac{1}{n}\left|\sum\limits_{x,y}\left(i_{x,y} - j_{x,y}\right)(-1)^{x+y}\right|, 
\end{equation*}

where $i_{x,y}$ is the orientation of the upper/left island in the horizontal/vertical pair located at coordinates $(x,y)$, and $j_{x,y}$ is the orientation of the lower/right island in the same pair (see Fig. 7b).

For the AFQ phase in which horizontal islands are fully polarized, we defined the AFQ order parameter as:

\begin{equation*}
\eta_{\mathrm{AFQ_x}}=\frac{4}{n^2}\left|\sum_{\substack{x,y \\ x+y\in\mathbb{E}}}i_{x,y} + j_{x,y}\right|\cdot\left|\sum_{\substack{x,y \\ x+y\in\mathbb{O}}}\left(i_{x,y} - j_{x,y}\right)(-1)^{x}\right|, 
\end{equation*}

where $\mathbb{E}$ and $\mathbb{O}$ are sets of even and odd integers, respectively, which means that the first sum runs over horizontal pairs of islands, while the second one runs over vertical pairs (see Fig. 7b). Analogously, we defined the AFQ order parameter for the AFQ phase with vertical islands fully polarized as:

\begin{equation*}
\eta_{\mathrm{AFQ_y}}=\frac{4}{n^2}\left|\sum_{\substack{x,y \\ x+y\in\mathbb{O}}}i_{x,y} + j_{x,y}\right|\cdot\left|\sum_{\substack{x,y \\ x+y\in\mathbb{E}}}\left(i_{x,y} - j_{x,y}\right)(-1)^{y}\right|, 
\end{equation*}

and so the global AFQ order parameter was defined as:

\begin{equation*}
\eta_{\mathrm{AFQ}}=\eta_{\mathrm{AFQ_x}} + \eta_{\mathrm{AFQ_y}.}
\end{equation*}

\begin{figure} 
\center
\includegraphics[width=.45\textwidth]{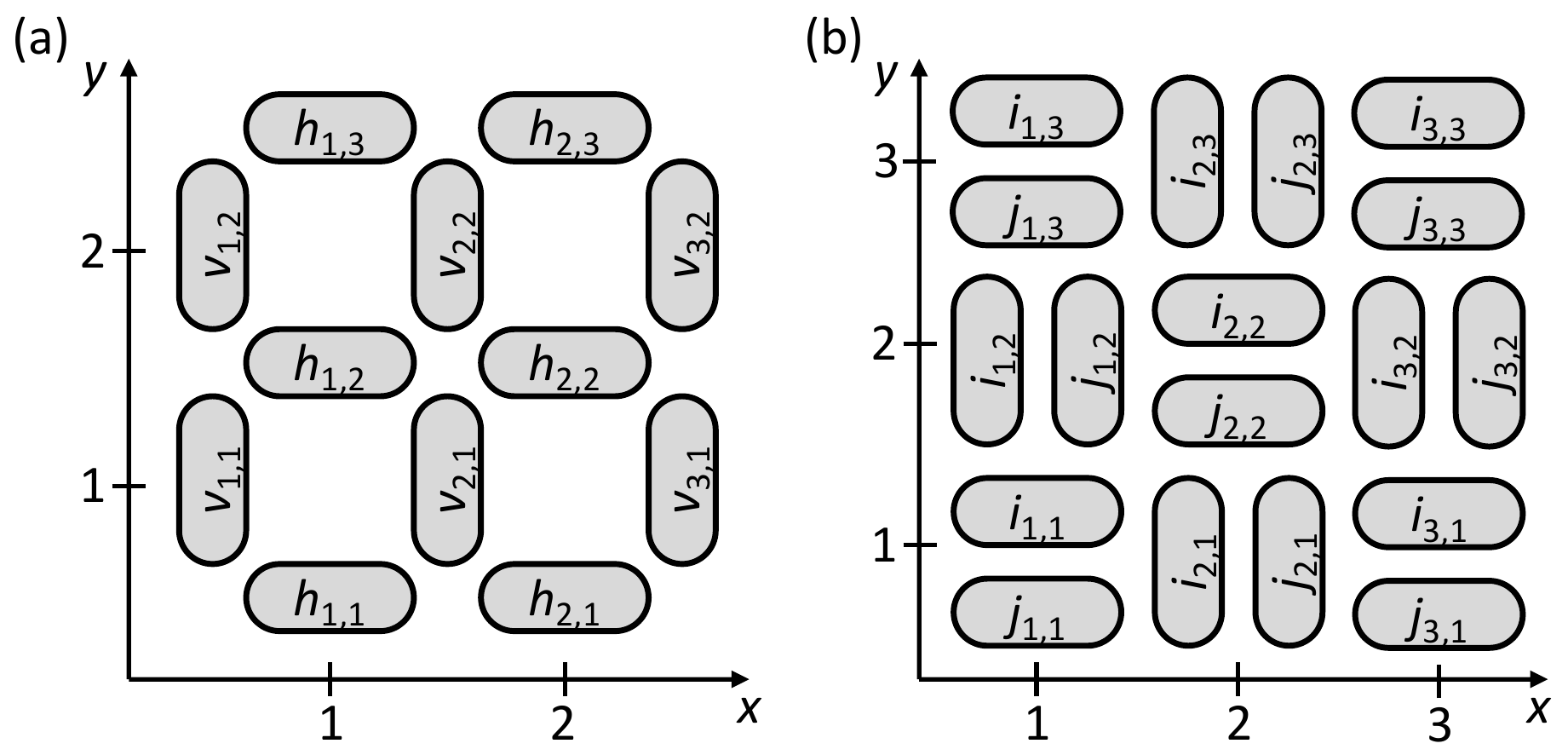}
\caption{Numbering of the islands in the a) square and b) quadrupolar lattices, using which the AF, FQ, and AFQ order parameters are calculated as described in the Appendix.}
\label{islandnumbering}
\end{figure}

\section{Acknowledgements} 

S.A.C, M.G., and C.N. gratefully acknowledge support from the Los Alamos LDRD program. The NHMFL is supported by the National Science Foundation (NSF) DMR-1644779, the State of Florida, and the US Department of Energy (DOE).  Permalloy growth (J.D.W. and C.L.) was supported by NSF DMR-1807124 and DMR-2103711. Sample design and patterning (X.Z. and P.S.) was funded by the US DOE, Office of Basic Energy Sciences, Materials Sciences and Engineering Division under Grant No. DE-SC0020162. Work at the University of Warsaw (M.G.) was supported by the Norwegian Financial Mechanism 2014-2021 under Grant No. 2020/37/K/ST3/03656 and from the Polish National Agency for Academic Exchange within Polish Returns program under Grant No. PPN/PPO/2020/1/00030. M.G. performed the noise measurements and the Monte Carlo simulations. S.A.C., M.G., and P.S. wrote the manuscript, with contributions from all authors. 


\end{document}